\documentclass[iop]{emulateapj}
\usepackage{color} 

\usepackage{amsmath}
\usepackage{amssymb}

\newcommand{\eqr}[1]{Eq.\thinspace(#1)}
\newcommand{\pfrac}[2]{\frac{\partial #1}{\partial #2}}

\newcommand{\mvec}[1]{\mathbf{#1}}

\newcommand{\gcs}{\nabla}
\newcommand{\gvs}{\nabla_{\mvec{v}}}

\newcommand{\gke}{\texttt{Gkeyll}}
\newcommand{\dHy}{\texttt{dHybridR}}

\begin{document}


\title{\sc{Temperature-dependent Saturation of Weibel-type Instabilities in Counter-streaming Plasmas}}


\title[Phase Space Energization of Ions in Oblique Shocks]{Phase Space Energization of Ions in Oblique Shocks}

\author{James Juno\thanks{Email address
for correspondence: jjuno@pppl.gov}$^{1}$, Collin~R.~Brown$^{2}$, Gregory~G.~Howes$^{2}$, Colby~C.~Haggerty$^{3}$, Jason~M.~TenBarge$^{4}$, Lynn B.~Wilson III$^{5}$, Damiano Caprioli$^{6}$, Kristopher G.~Klein$^{7}$}
\affiliation{$^1$Princeton Plasma Physics Laboratory, Princeton, NJ 08540, USA\\
$^2$Department of Physics and Astronomy,University of Iowa, Iowa City IA 52242, USA\\
$^3$Institute for Astronomy, University of Hawaii, Honolulu, HI 96822, USA\\
$^4$Department of Astrophysical Sciences, Princeton University, Princeton, NJ 08544, USA\\
$^5$ NASA Goddard Space Flight Center, Heliophysics Division, Greenbelt, MD 20771, USA\\
$^6$ Department of Astronomy and Astrophysics, University of Chicago, Chicago, Illinois 60637, USA\\
$^7$ Lunar and Planetary Laboratory, University of Arizona, Tucson, AZ 85719, USA\\
}


\begin{abstract}
Examining energization of kinetic plasmas in phase space is a growing topic of interest, owing to the wealth of data in phase space compared to traditional bulk energization diagnostics.
Via the field-particle correlation (FPC) technique and using multiple means of numerically integrating the plasma kinetic equation, we have studied the energization of ions in phase space within oblique collisionless shocks.
The perspective afforded to us with this analysis in phase space allows us to characterize distinct populations of energized ions.
In particular, we focus on ions which reflect multiple times off the shock front through shock-drift acceleration, and how to distinguish these different reflected populations in phase space using the FPC technique.
We further extend our analysis to simulations of three-dimensional shocks undergoing more complicated dynamics, such as shock ripple, to demonstrate the ability to recover the phase space signatures of this energization process in a more general system.
This work thus extends previous applications of the FPC technique to more realistic collisionless shock environments, providing stronger evidence of the technique's utility for simulation, laboratory, and spacecraft analysis.

\end{abstract}




\section{Introduction}
The means by which the hot, tenuous plasmas that make up the luminous universe convert energy from one form to another via collisionless processes, i.e., processes involving the interaction between the constituent plasma particles and the self-consistently generated electromagnetic fields, remains an active area of research.
In particular, the mechanisms for dissipation of energy in collisionless shock-waves are a rich field of study.
Here, a shock-wave refers to a disturbance propagating faster than the largest local wave speed, and the shock-wave is considered collisionless because the steepening of the shock, and the resultant conversion of the upstream kinetic energy to other forms of energy, occurs on a length scale much smaller than the collisional mean free path of the plasma particles.
How the plasma converts the large bulk kinetic energy of the incoming supersonic flow to electromagnetic and thermal energy over such small length scales can vary dramatically based on the shock geometry and the fast magnetosonic Mach number $M_{f} = U_{shock}/v_{f}$, where $U_{shock}$ is the shock velocity and $v_{f}$ is the fast magnetosonic wave velocity.
A critical commonality is that because these processes are collisionless, they involve the evolution of the plasma in the full position-velocity six-dimensional (3D-3V) phase space.

Leveraging phase space to determine the details of the plasma's evolution has historically been both fruitful but also fraught with challenges. 
Sufficient particle statistics to reconstruct the distribution function in both spacecraft observations and particle-in-cell (PIC) simulations is non-trivial and often limited by instrumental and computational constraints.
Nevertheless, the increasing resolution of modern spacecraft observations and numerical methods, along with more advanced laboratory measurement techniques for reconstructing distribution functions, is building towards ever-more sophisticated means of diagnosing the energetics of collisionless plasma phenomena such as collisionless shocks.

In particular, recent work has focused on the field-particle correlation technique \citep{Klein:2016,Howes:2017,Klein:2017a} to characterize the energy transfer between the plasma particles and electromagnetic fields in phase space.
First developed for turbulence applications \citep{Klein:2017b,Howes:2018,Li:2019,Klein:2020,Horvath:2020}, the FPC technique has recently been applied to collisionless shocks \citep{Juno:2021} to identify the energization signatures of shock-drift acceleration of ions \citep{Paschmann:1982, Sckopke:1983, Ball:2001} and adiabatic heating of electrons \citep[see e.g.,][]{Balogh:2013} in phase space.
These velocity-space signatures of the exchange of energy between particles and electromagnetic fields are useful for understanding energization in kinetic simulations and of especially high utility when comparing to laboratory measurements\citep{Schroeder:2021} and spacecraft observations \citep{Chen:2019,Afshari:2021}.
Spacecraft, for example, provide us an Eulerian perspective of how regions of phase space are being energized after sufficient integration over particle statistics, in contrast to a Lagrangian perspective of tracking the energization of individual particles.
There is thus growing interest in characterizing a wider variety of collisionless shock energization mechanisms in phase space using the field-particle correlation technique and comparing the observed signatures from spacecraft measurements of distribution functions in heliospheric shocks to those obtained from simulations.

\citet{Juno:2021} considered a fairly idealized shock: one-dimensional and perpendicular, where perpendicular defines the angle between the upstream magnetic field and shock-normal direction.
In more realistic plasma environments, not only is there often a finite component of the magnetic field in the shock-normal direction, but also instabilities in the transverse plane of the shock \citep{Schwartz:1985, Schwartz:1992, Wilson:2013b, Wilson:2014b}, which may lead to distortions and a corrugation of the shock, often referred to as ``shock ripple'' \citep{Johlander:2016}.
In both cases, this additional physics may significantly complicate the shock dynamics, allowing particles to return upstream when the magnetic field partially lies in the shock-normal direction.
Further, this additional physics may prevent a clean ``two-state'' picture of the shock, where there is a clearly defined upstream and downstream region with respect to the shock, because these transverse instabilities distort the shock transition.
It is thus reasonable to ask whether these potentially powerful velocity-space signatures of energization in phase space are recoverable in these more realistic shock environments.

It is the purpose of this manuscript to address this issue with both more realistic shock geometries and full three-dimensional simulations which permit the development of transverse instabilities.
We again leverage the \gke\ simulation framework \citep{Juno:2018,HakimJuno:2020} to perform continuum kinetic simulations of a collisionless shock and obtain high fidelity representations of the distribution function through the shock, free of the counting noise present in particle-in-cell simulations.
Similar to \citet{Juno:2021}, we perform one dimensional simulations of an electron-ion plasma but now equally subdivide the magnetic field between the components perpendicular and parallel to the shock-normal direction, often referred to as an oblique shock with shock-normal angle $\theta_{Bn} = 45^{\circ}$.
We find with these one dimensional simulations of an oblique shock that not only can we still recover the shock-drift acceleration signature in this more realistic shock geometry, but the inclusion of an in-plane magnetic field component permits the ions to bounce multiple times off the shock front and thus be continually energized by shock-drift acceleration.
Using the field-particle correlation technique, we can clearly identify these distinct populations of energized ions.

We supplement this more computationally demanding model with complementary hybrid particle-in-cell simulations using \dHy \citep{Gargate:2007,Haggerty:2019a,Haggerty:2020,Caprioli:2020} in three dimensions with the same initial shock geometry.
We are able to obtain identical velocity-space signatures of the ions bouncing multiple times off the shock front and being energized continually via shock-drift acceleration even with the added complication of the shock distorting and rippling in the transverse plane due to a kinetic instability.
This study thus serves also as a proof-of-concept that this phase-space analysis can also be performed on particle-in-cell simulations, provided a sufficiently large number of particles is used to allow for a reasonably accurate reconstruction of the distribution function.

\section{Computational Models and the Field-Particle Correlation Technique}\label{sec:compModel}

Both the continuum Vlasov-Maxwell solver in the \gke~framework and the hybrid particle-in-cell (PIC) code \dHy\ numerically integrate the Vlasov equation,
\begin{align}
    \pfrac{f_s}{t} + \gcs \cdot (\mvec{v} f_s) + \gvs \cdot \left [ \frac{q_s}{m_s}(\mvec{E} + \mvec{v} \times \mvec{B}) f_s \right ] = 0, \label{eq:vlasov}
\end{align}
where $f_s = f_s(\mvec{x}, \mvec{v}, t)$ is the particle distribution function for species $s$, $q_s$ and $m_s$ are the charge and mass of species $s$ respectively, and $\mvec{E} = \mvec{E}(\mvec{x},t)$ and $\mvec{B} = \mvec{B}(\mvec{x},t)$ are the electric and magnetic fields respectively.
We note that collisions can be included in the \gke\ discretization of the Vlasov equation as a Fokker-Planck operator \citep{Hakim:2020} on the right-hand-side of \eqr{\ref{eq:vlasov}}, and that such an operator is useful for providing numerical regularization of velocity space given finite velocity space resolution. 
The two methods utilize different field equations for the evolution of the electromagnetic fields; \gke's Vlasov-Maxwell solver numerically integrates Maxwell's equations, while \dHy\ uses a reduced set of field equations in the limit that the electron mass is negligible.

Both methods have advantages and disadvantages for this particular study.
\gke's continuum Vlasov-Maxwell solver directly discretizes the Vlasov equation on a phase space grid, thus completely eliminating the noise inherent in the PIC algorithm which can be problematic for distribution function analysis.
Discretization on a phase space grid comes at an increased computational cost---the one dimensional oblique shock analyzed in Section~\ref{sec:1D} requires a four dimensional phase space (1D-3V).
\dHy's cost is further decreased by its reduced electromagnetic field equations which allow \dHy\ to step over restrictive electron temporal and spatial scales.
While the approximate electromagnetic field equations prevent \dHy\ from being used to examine electron energization, the reduced model does lower the computational cost enough for us to perform three-dimensional simulations in configuration space of the ion dynamics.
Additionally, the lowered computational cost allows us to utilize a large particle-per-cell count and thus increase the accuracy of the distribution function reconstruction by decreasing the noise caused by the sampling of velocity space using discrete particles.

With access to the distribution function, we can employ the foundational tool of this study: the field-particle correlation (FPC) technique.
By defining the phase space energy density $w_s(\mvec{x},\mvec{v},t) \equiv m_s v^2 f_s(\mvec{x},\mvec{v},t)/2$ and multiplying the Vlasov equation by $m_s v^2/2$, we obtain an expression for the rate of change of this phase-space energy density,
\begin{align}
  \frac{\partial w_s}{\partial t} = - \mvec{v}\cdot \nabla  w_s  -
  q_s\frac{v^2}{2}  \mathbf{E} \cdot \frac{\partial f_s}{\partial \mathbf{v}}
   - q_s\frac{v^2}{2} \left(\mathbf{v} \times \mathbf{B}\right)
      \cdot \frac{\partial f_s}{\partial \mathbf{v}}.
  \label{eq:dws} 
\end{align}
The only term which leads to net energy transfer in phase space is the electric field term \citep{Klein:2016, Howes:2017, Klein:2017b}, which allows us to define
\begin{align}
    C_\mvec{E} & (\mvec{x}_0,\mvec{v},t;\tau) = \notag \\
    & \frac{1}{\tau} \int_{t-\tau/2}^{t+\tau/2} \left[-q_s\frac{v^2}{2} 
    \frac{\partial f_s(\mvec{x}_0,\mvec{v},t')}{\partial \mvec{v}} \right] \cdot \mvec{E}(\mvec{x}_0,t')\thinspace  dt'. \label{eq:FPC-full}
\end{align}
\eqr{\ref{eq:FPC-full}} is the principal instrument for our subsequent analysis and provides a measure of the rate of change of phase-space energy density at position $\mvec{x}_0$ as a function of velocity space $\mvec{v}$ over the correlation time $\tau$.
We call the resulting signature the \emph{velocity-space signature} characteristic of the mechanism of energization.
Different velocity-space signatures can then be used to identify a particular energization process such as a wave-particle resonance like Landau damping \citep{Howes:2017, Klein:2017b} and cyclotron damping \citep{Klein:2020}, or in the case of the previous \citet{Juno:2021} shock study, a non-resonant signature such as shock-drift acceleration or adiabatic heating.


For the present study of ion energization at collisionless shocks using the FPC technique, we follow \citet{Juno:2021} to separate the correlation by electric field component $E_j$ and to take $\tau=0$ so that our correlation is instantaneous\footnote{In the presence of shock non-stationarity, such as shock reformation, a finite correlation interval is likely required to average over the reformation cycle, see, e.g, \citet{Balogh:2013, Caprioli:2014d} for discussions of shock non-stationarity. Over the time interval of analysis, no shock non-stationarity is observed for the simulations presented in this study.}. 
Here we emphasize two key concepts in the implementation of the FPC technique:  (i) the choice of reference frame and (ii) the different means of calculating the correlation for a grid code versus a particle code.

First, the velocity-space signatures of ion energization are most easily interpreted if the calculations are performed in a frame of reference in which the upstream flow is along the shock normal (normal incidence frame) and in which the shock is at rest (shock-rest frame). 
The upstream inflow is easily initialized along the shock normal; see Appendix~\ref{app:gke-param} for details.  
In the frame of the simulations (the downstream rest frame), the shock propagates at velocity $\mvec{U}_{shock}$, so the electromagnetic fields are Lorentz transformed, in the non-relativistic limit, from the simulation frame (primed) to the shock-rest frame (unprimed) by 
\begin{align}
    \mvec{E} & = \mvec{E}' + \mvec{U}_{shock} \times \mvec{B}', \\
    \mvec{B} & = \mvec{B}',
\end{align}
and velocity coordinates are likewise shifted by $\mvec{v}=\mvec{v}'-\mvec{U}_{shock}$. 
In the resulting shock-rest frame, the contribution to the ion energization due to the $j$th component of the electric field is therefore given by
\begin{align}
    C_{E_j} (x_0, \mvec{v}, t) & = -q_i \frac{v_j^2}{2} E_j(x_0,t) \pfrac{f_i(x_0, \mvec{v}, t)}{v_j}, \label{eq:vjCorrelation}
\end{align}

Second, although for the \gke\ simulations we can directly compute \eqr{\ref{eq:vjCorrelation}} from the simulation output because we have the distribution function at every grid point in configuration space, for the \dHy\ simulations, integrating over a finite spatial volume is required to reconstruct the distribution function from the particles.
If we bin the particles to obtain $f_i$ before computing the correlation, the electric field used in the correlation must be spatially averaged over the binning volume.
Instead, following \citet{Chen:2019}, we compute for every particle an alternative correlation,
\begin{align}
    C'_{E_j}(\mvec{x}_l, \mvec{v}_l, t) = q_i v_{j_l} E_j(\mvec{x}_l), \label{eq:Cprime}
\end{align}
where $l$ denotes the $l^{\textrm{th}}$ ion.
We then bin \eqr{\ref{eq:Cprime}} into equally spaced bins in configuration and velocity space.
This pre-computation of the correlation before binning allows us to retain the spatial variation of the electric field within the binning volume, reducing noise and improving accuracy.
We then recover the original definition of the component-wise correlation with
\begin{align}
    C_{E_j}(\mvec{x}_0, \mvec{v}, t) = -\frac{v_j}{2} & \pfrac{}{v_j} \left [ \sum_l C'_{E_j}(\mvec{x}_l, \mvec{v}_l, t) \Delta \mvec{x} \Delta \mvec{v} \right ] \notag \\
    & + \frac{1}{2} \left [ \sum_l C'_{E_j}(\mvec{x}_l,  \mvec{v}_l, t) \Delta \mvec{x} \Delta \mvec{v} \right ], \label{eq:ptcl-fpc}
\end{align}
where $\Delta \mvec{x} \Delta \mvec{v}$ are the sizes of our configuration space and velocity space bins, and the sum over $l$ sums over all the ions in that bin.

Finally, we note that we generally integrate the computed correlations, \eqr{\ref{eq:vjCorrelation}}, over one velocity degree of freedom for ease of visualization.
In other words, reduced quantities such as
\begin{align}
    C_{E_y}(x_0, v_x, v_z, t) & \equiv \int C_{E_y}(x_0, \mvec{v}, t) \thinspace dv_y, \notag \\
    & = \int -q_i \frac{v_y^2}{2} E_y(x_0, t) \pfrac{f_i(x_0, \mvec{v}, t)}{v_y} \thinspace dv_y, 
\end{align}
facilitate 2V visualizations of velocity-space signatures, thereby avoiding the complexities of three-dimensional visualization. 
Henceforth, the velocity-space dependence of $C_{E_j}$ will be explicitly stated, with any missing velocity coordinates implying integration over that dimension.

\section{Field-Particle Correlation Analysis of Ions: 1D Oblique Shock in \gke}\label{sec:1D}

To simulate an oblique collisionless shock self-consistently using the continuum Vlasov-Maxwell solver in \gke, we set up a one-dimensional geometry in configuration space with all three velocity dimensions (1D-3V) and choose the one spatial coordinate to be along the shock normal in the $x$ direction.
The initial magnetic field is then taken to be in the $x-z$ plane, $\mvec{B} (t = 0) = (B_0 \mvec{\hat{x}} + B_0 \mvec{\hat{z}})/\sqrt{2}$.
We choose a domain size of $L_x = 24 d_i$ and an in-flow velocity to initiate the shock of $U_x = 6 v_A$, where $d_i = c/\omega_{pi}$ is the ion inertial length and $v_A = B_0/\sqrt{\mu_0 n_0 m_i}$ is the ion Alfv\'en speed respectively.
Here, $c$ is the speed of light, $m_i$ is the ion mass, $\omega_{pi} = \sqrt{e^2 n_0/\epsilon_0 m_i}$ is the ion plasma frequency, and the subscript $0$ denotes the upstream value, e.g., $n_0$ is the upstream density and $B_0$ is the upstream magnetic field magnitude.
We employ a reduced mass ratio $m_i/m_e = 100$ where $m_e$ is the electron mass; other parameters are detailed in Appendix~\ref{app:gke-param}.

\begin{figure}
    \centering
    \includegraphics[width=\linewidth]{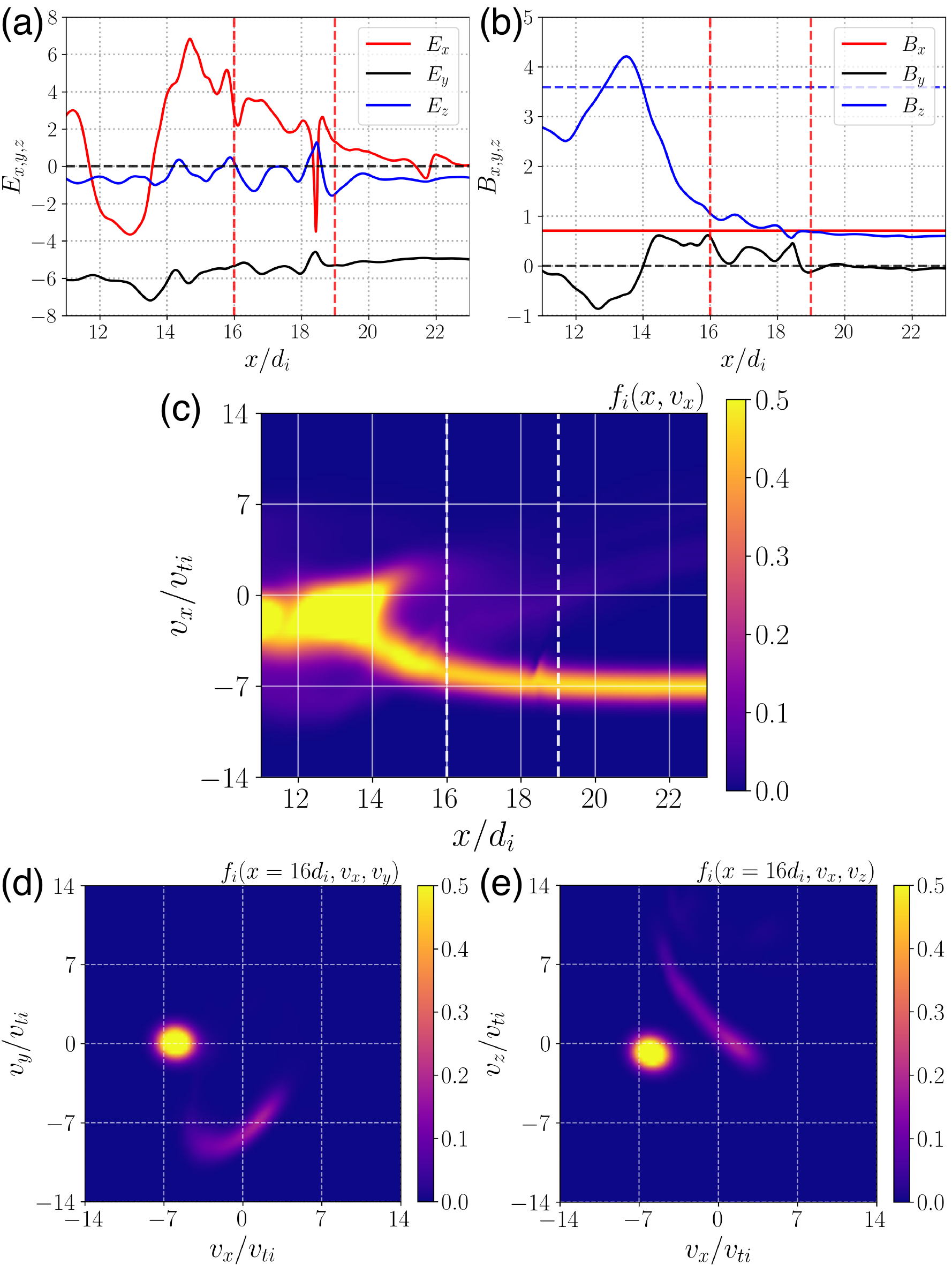}
    \caption{Electromagnetic fields (a) and (b), ion distribution function as a function of $(x, v_x)$ (c), and ion distribution function at $x = 16 d_i$ in the shock ramp as a function of $(v_x, v_y)$ (d) and $(v_x, v_z)$ (e). In the shock-rest frame, $E_y$ and $E_z$ are approximately constant through the shock, while the cross shock electric field $E_x$ arises to maintain quasi-neutrality as a result of the electron pressure gradient. $B_z$ compresses as a result of the shock, while $B_x$ remains constant in this one dimensional geometry because $\nabla \cdot \mvec{B} = 0$. In all three means of visualizing the ion distribution function, we observe a reflected ion population as a result of ions upstream encountering the compressed magnetic field and returning upstream due to their gyromotion.}\label{fig:gke-shock}
\end{figure}

We visualize the electromagnetic fields along with the corresponding ion distribution functions in the shock-rest frame in Figure~\ref{fig:gke-shock} at $t = 8 \Omega_{ci}^{-1}$.
We restrict our attention to just the region immediately in proximity to the shock: the foot, ramp, overshoot, and transition to the downstream.
We mark an approximate compression ratio of $r = 3.6$ of the compressing component of the magnetic field, $B_z$, which we have obtained from the MHD Rankine-Hugoniot solutions for a shock with the parameters: $\theta_{Bn} = 45^{\circ}, \beta_{tot} = 2, M_A = 8.3$.
This compression ratio corresponds to a shock velocity of $\mvec{U}_{shock} = U_x/(r-1) \hat{\mvec{x}} \sim 2.3 v_A \hat{\mvec{x}}$ for transforming the electromagnetic fields to the shock rest frame and a final Alfv\'en Mach number of $M_A \sim 8.3$ in the shock-rest frame.
We focus in particular on the ion distribution function in the $v_x-v_y$ and $v_x-v_z$ planes at $x = 16 d_i$ and $x = 19 d_i$ in the ramp and foot of the shock respectively for further analysis.
We find a crescent-shaped structure at $x = 16 d_i$ in the  $v_x-v_y$ plane for the reflected ion population, similar to the structure observed in the perpendicular shock study in \citet{Juno:2021} and other simulation studies of shock-drift acceleration \citep[e.g.,][]{Park:2013, XGuo:2014a, XGuo:2014b, Park:2015, Xu:2020}.

\begin{figure*}
    \centering
    \includegraphics[width=\textwidth]{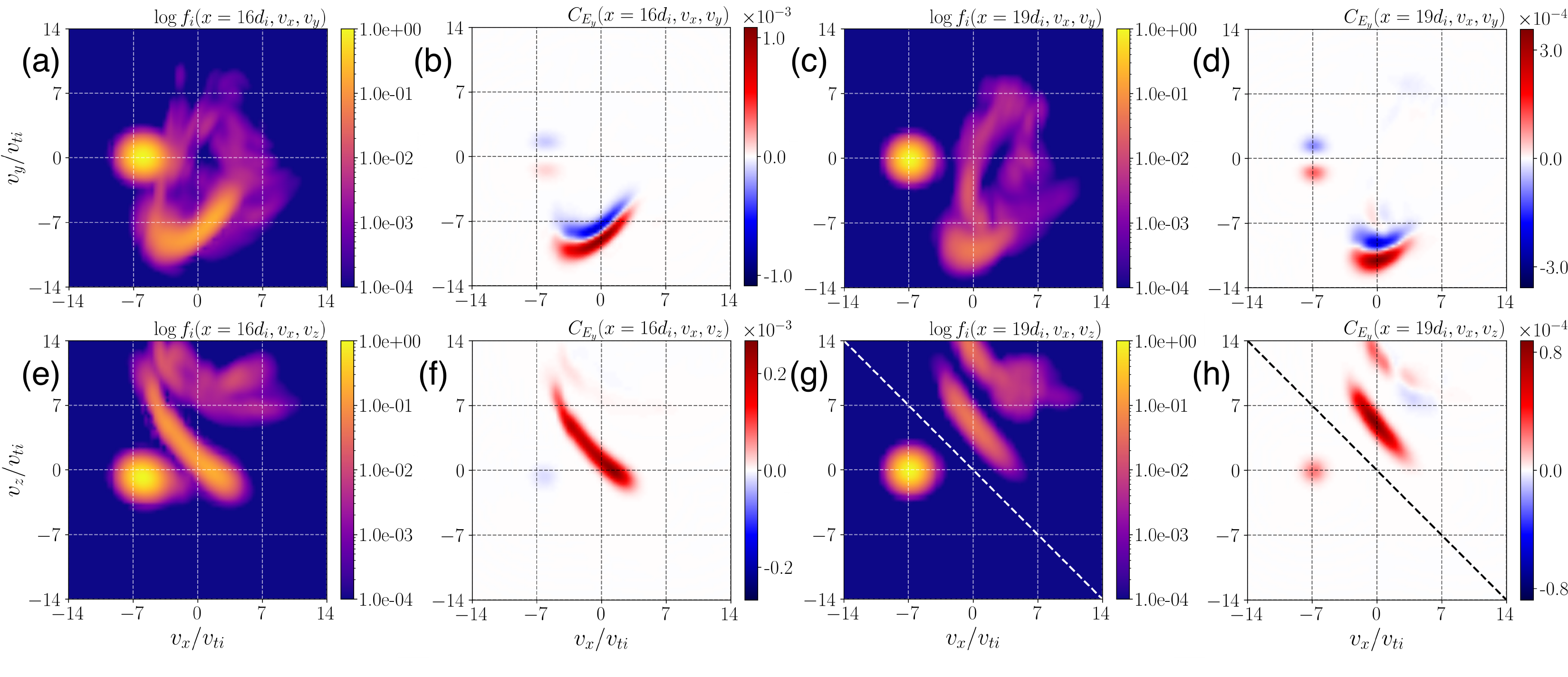}
    \caption{Ion distribution functions and $C_{E_y}$ field-particle correlations from the \gke\ simulation at two locations: the shock foot at $x = 16 d_i$ and the shock ramp at $x = 19 d_i$. We plot the distribution function and correlations as a function of $v_x-v_y$, (a)--(d), and as a function of $v_x-v_z$, (e)--(h). A unique feature of this study with a more general shock geometry is the distribution function and correlations in the coplanarity plane $v_x-v_z$. While we note in both the shock foot and shock ramp we observe the same reflected population and velocity-space signature of shock-drift acceleration found in \citet{Juno:2021}, the Eulerian perspective of the distribution function in the coplanarity plane reveals additional insights. Firstly, the shock-drift acceleration signature appears as a positive-definite velocity space signature in this plane, owing to the integration over $v_y$ so that we identify all particles in phase space with particular $v_x$ and $v_z$ being energized. Secondly we can identify distinct populations of energized ions for particles that have bounced multiple times off the shock front, i.e., particles at $x = 19 d_i$, $v_x \sim [-3.5 v_{ti}, 3.5 v_{ti}], v_z \sim [0, 7 v_{ti}]$, which have undergone their first bounce and $v_x \sim [-1.0 v_{ti}, 1.0 v_{ti}], v_z \sim [10, 14 v_{ti}]$ which have undergone a second bounce.}\label{fig:gke-corr}
\end{figure*}

Motivated by the results of \citet{Juno:2021}, we plot the $C_{E_y}$ field-particle correlation in $v_x - v_y$ and $v_x - v_z$ at these two spatial locations in Figure~\ref{fig:gke-corr}.
To accentuate the structure in the distribution function, we plot the logarithm (base 10) of the distribution function.
In fact, from this perspective, not only do we see the familiar reflected ion population in the $v_x-v_y$ plane, but now we can identify a separate, higher energy population in the $v_x-v_z$ plane.
Note that the $v_x-v_z$ plane corresponds to the velocities in the upstream coplanarity plane.

Further, in the $v_x-v_y$ plane at both $x = 16 d_i$ in the ramp and $x = 19 d_i$ in the foot, we observe the same velocity-space signature for shock-drift acceleration: a blue-red signature coincident with the crescent feature in the ion distribution function corresponding to a loss (blue) of phase space energy density at lower energies and a gain (red) of phase space energy density at higher energies.
The crescent-shaped feature in phase space represents the reflected ions resulting from the ion's gyromotion being turned around by a combination of the shock's magnetic field gradient and, to a lesser extent, the cross-shock electric field.
These ions pass back upstream into a region of lower magnetic field, and can then gain energy from the motional electric field which supports the incoming supersonic $\mvec{E} \times \mvec{B}$ flow.
This same velocity-space signature can be viewed from a different perspective in the $v_x-v_z$ plane, where all particles in the range of velocities\footnote{Here, $v_{ti} = \sqrt{2 k_B T_{i_0}/m_i}$ is the upstream ion thermal velocity and $k_B$ and $T_{i_0}$ are Boltzmann's constant and the upstream ion temperature respectively. See Appendix~\ref{app:gke-param} for a further discussion of the velocity space grid employed for these simulations.} $v_x \sim [-3.5 v_{ti}, 3.5 v_{ti}], v_z \sim [0, 7 v_{ti}]$ are particles which have reflected off the shock front and can thus gain energy via shock-drift acceleration.
In fact, there is an additional small signal in the velocity space signature coincident with the higher energy population population at $x = 16 d_i$ in the ramp, and the signal becomes stronger as we move into the foot of the shock, $x = 19 d_i$.

Thus, not only can we easily obtain the shock-drift acceleration velocity-space signature found in \citet{Juno:2021} in this more general shock geometry, we can also obtain an energization signature for a separate, higher energy ion population.
This additional velocity space signature corresponds to ions which have experienced an additional bounce off the shock, i.e., another round of shock-drift acceleration.
Because these now higher energy particles have a larger Larmor radius and thus gyrate further upstream when they gain energy, the signature is stronger in the foot of the shock.
In fact, for this shock geometry and Mach number, many of these particles that are being further energized upstream of the shock return upstream instead of passing downstream after their second bounce--see Appendix \ref{app:spm} for a similar analysis to \citet{Juno:2021} connecting the single-particle-motion picture to the observed distribution functions in phase space. 

Further, with access to all of phase space we can also quantitatively assess the energization of the particles in ways which are impossible with traditional bulk energization diagnostics, such as $\mvec{J} \cdot \mvec{E}$.
We have demarcated in panels (g) and (h) of Figure \ref{fig:gke-corr} a separation between the parts of phase space which contain the incoming supersonic beam and the reflected particles, $v_x = - v_z$.
Integrating $C_{E_y}$ over the remaining velocity dimensions in phase space below and above this line separates the energization of the incoming beam and reflected particles.
We find that, while the beam density is 5.22 times greater than the reflected particle density, the reflected particles experience a 10.1 times larger increase in their energy density.
Thus, per particle, the reflected particles gain 52.7 times more energy than the energy change experienced by the incoming beam of supersonic particles.
We emphasize again that this type of analysis is not possible without access to phase space; by examining phase space holistically, we gain new perspective on how ions across energy scales are heated and accelerated.


\section{Field-Particle Correlation Analysis of Ions: 3D Oblique Shock in \dHy}\label{sec:3D}

While the shock-drift acceleration velocity-space signature has been recovered in a more general shock geometry in one spatial dimension, and even revealed further benefits of the field-particle correlation approach through the identification of an additional velocity-space signature for ions bouncing multiple times off the shock, full three-dimensional simulations allow for even further complications to the shock dynamics.
For the \dHy\ three-dimensional shock, we adopt a similar shock geometry to the \gke\ simulation analyzed in Section~\ref{sec:1D}: the shock normal is in the $x$ direction and the initial magnetic field is in the $x-z$ plane as before.
We use a standard set-up employed in previous hybrid PIC shock simulations \citep{Gargate:2007,Haggerty:2019a} where particles are injected from one end of the domain ($x = L_x$) and reflect off a conducting wall at the other end of the domain ($x = 0$) to drive a shock which propagates away from this reflecting wall in the simulation frame.
The in-flow velocity of the particles is identical to the \gke\ simulation, $U_x = - 6 v_A$.
Periodic boundary conditions are employed in $y$ and $z$, and further parameters can be found in Appendix~\ref{app:dhy-param}.

\begin{figure}
    \centering
    \includegraphics[width=\linewidth]{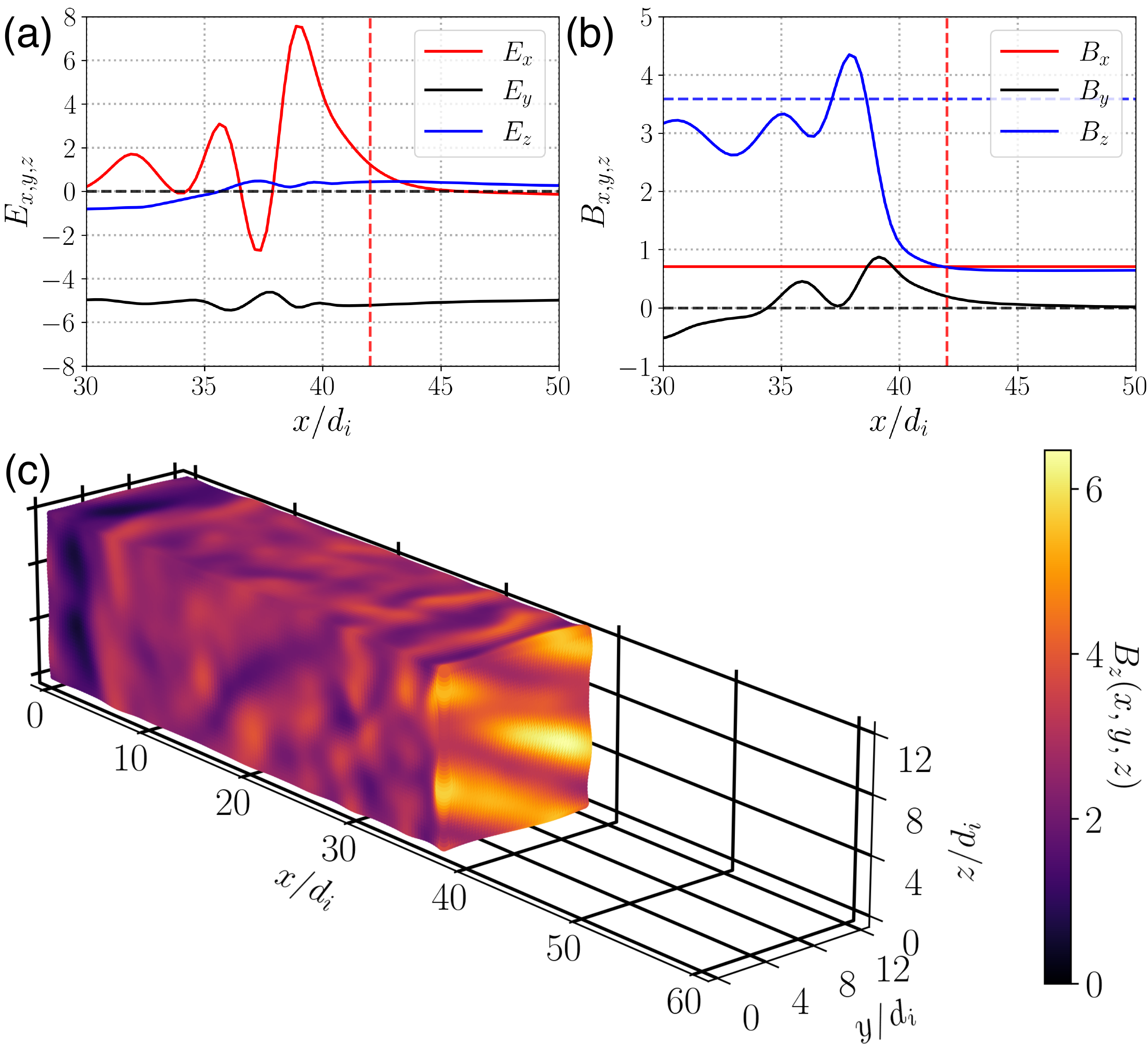}
    \caption{Electromagnetic fields in the shock-rest frame integrated over the transverse directions plotted in the shock-normal direction, (a) and (b), and three dimensional isosurface of $B_z$ (c) from the \dHy\ simulation. The one-dimensional profiles are similar to the one-dimensional \gke\ simulations, but the three-dimensional visualization demonstrates the additional physics present in the higher dimensional \dHy\ simulation: instabilities in the transverse plane lead to a rippling of the shock such that the shock is not completely planar. We mark an approximate compression ratio of the shock $r \sim 3.6$ and the $x$ location of interest in the shock foot for the subsequent distribution function analysis, $x = 42 d_i$.}\label{fig:dhy-shock}
\end{figure}

\begin{figure*}
    \centering
    \includegraphics[width=\textwidth]{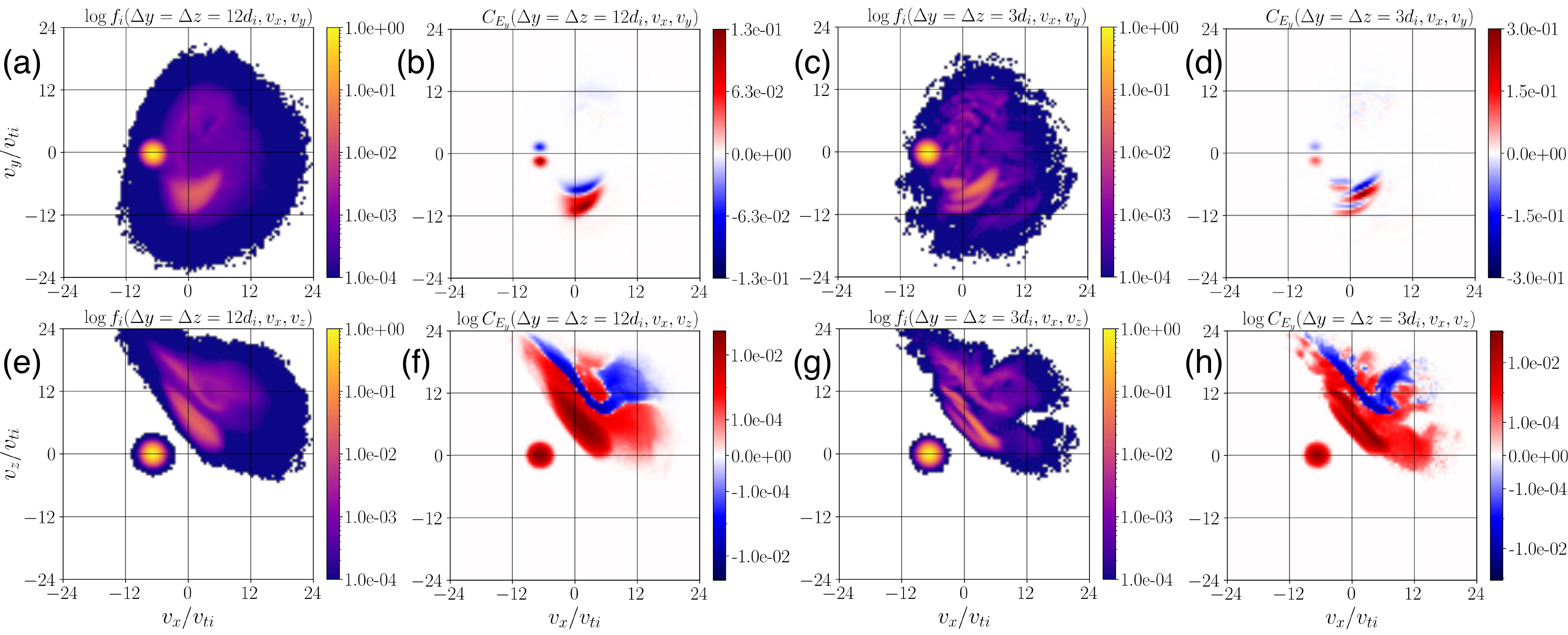}
    \caption{Ion distribution functions and $C_{E_y}$ field-particle correlation from the three-dimensional \dHy\ simulation, averaged over $\Delta x = d_i/2$, centered at $x = 42 d_i$ in the shock foot for two transverse averaging boxes: one, the entire transverse direction, $\Delta y = \Delta z = 12 d_i$, of the three-dimensional simulation (a),(b),(e), and (f), and two, a smaller region of the transverse direction, $\Delta y = \Delta z = 3 d_i$, from $y = [0, 3 d_i], z = [0, 3 d_i]$ (c),(d),(g), and (h). Note that we plot the $C_{E_y}$ field-particle correlation in the coplanarity plane, $v_x-v_z$, with a symmetric logarithmic colorbar to more easily identify the particles which have bounced twice off the shock front. While the need to average over a finite box extent in a PIC method reduces the solution quality compared to a point-wise distribution function obtained from the Eulerian \gke\ simulation and averages over different regions of the shock due to the shock ripple, mixing upstream, foot, ramp, and overshoot particles, we nonetheless can obtain the velocity-space signature of shock-drift acceleration and identify the same multiple bounce particles in phase space.}\label{fig:dhy-corr}
\end{figure*}

We plot the electromagnetic fields from the \dHy\ simulation in Figure \ref{fig:dhy-shock}, averaging $\mvec{E}$ and $\mvec{B}$ over the transverse dimensions for comparison with the \gke\ electromagnetic fields in Figure~\ref{fig:gke-shock}, and the compressing component of the magnetic field, $B_z$, in three dimensions to illustrate the non-trivial structure along the shock interface.
This corrugation or rippling of the shock is the product of instabilities driven by the reflected ion population and can further modify the shock energetics.
We reconstruct the distribution function and compute the FPC using \eqr{\ref{eq:ptcl-fpc}} for two different transverse spatial integration areas in Figure \ref{fig:dhy-corr} and a fixed shock-normal direction integration range $x = [41.75 d_i, 42.25 d_i]$, i.e., $\Delta x = d_i/2$ centered at $x = 42 d_i$.

Integrating over the entire transverse direction, $\Delta y = \Delta z = 12 d_i$, allows us to construct a representation of the distribution function with little apparent particle noise, and we recover both the familiar blue-red crescent velocity-space signature of shock-drift acceleration in $v_x-v_y$ and can identify the multiple-bounce velocity-space signature in $v_x-v_z$.
However, such a large integration window averages over this shock ripple and broadens some of the features of the velocity-space signatures, as different regions of the transverse plane may be in different parts of the shock, i.e., the foot where the second ion bounce is more easily visible versus the ramp where the first ion bounce is more strongly energized.
Integrating over a smaller region of the transverse direction, $\Delta y = \Delta z = 3 d_i$, from $y = [0, 3 d_i], z = [0, 3 d_i]$, reduces the quality of the distribution function reconstruction, but nevertheless still allows for a reasonable computation of the FPC.
Thus, despite the shock ripple complicating the shock transition, we can still safely apply the FPC to understand the energization of particles in phase space.
Further, while the grid-based solution provided by \gke\ is of high utility for this type of phase-space analysis, with sufficiently high particle counts and a reformulation of the FPC for particle data provided by \eqr{\ref{eq:ptcl-fpc}}, we can diagnose the energization of the plasma in phase space with a PIC method.

\section{Summary and Future Outlook}\label{sec:conclusions}
Motivated by the growing interest in leveraging phase space to understand particle energization in collisionless shocks, this study has three principal results:
\begin{itemize}
    \item The velocity-space signature of shock-drift acceleration found using the field-particle correlation technique in \citet{Juno:2021} for an idealized perpendicular shock can be found in more realistic shock geometries where the upstream magnetic field is almost never completely perpendicular to the shock-normal direction.
    \item The velocity-space signature of ions which have bounced multiple times off the shock, undergoing multiple rounds of shock-drift acceleration, has likewise been found, and further demonstrates the utility of the field-particle correlation technique, as the technique provides a holistic view of phase space in which distinct populations of energized ions can be identified and examined.
    \item This holistic view of phase space allows for unique quantitative analysis of phase space by permitting separate calculations of the energization experienced by different populations of particles.
    The energy gain per particle of the reflected populations is over 50 times larger than the energy change per particle of the incoming beam, a determination we can make because we can sub-divide phase space and compute the energization of each population of particles.
    \item These signatures persist into general three-dimensional simulations where the shock distorts in the transverse plane due to kinetic instabilities, and further with sufficient particle resolution, these analysis techniques are applicable to traditional particle-in-cell methods for the simulation of collisionless shocks.
\end{itemize}
Importantly, this phase-space analysis procedure generalizes far beyond the cases considered in this study.
As these shock waves propagate for greater distances, particles may undergo even further energization, reflecting continuously both off the shock and off electromagnetic fluctuations upstream driven unstable by the reflected particles in the classical diffusive shock acceleration picture of energization \citep{Fermi:1949, Fermi:1954, Blandford:1978, Caprioli:2010b}.
The field-particle correlation technique not only permits a visualization in phase space of each distinct population of particles' energetics, but also allows for a fine-grained quantitative analysis in which we can integrate over sub-domains of phase space to understand both how these particles are energized and the degree to which they are energized.
Phase space analysis thus provides benefits compared to both the tracking of the energetics of individual particles and to traditional bulk energization diagnostics such as $\mvec{J} \cdot \mvec{E}$.

We defer to future work a systematic convergence study of the minimum particle resolution to reconstruct distribution functions to the level of fidelity required to compute the field-particle correlation.
Such a study requires a quantitative criterion for when a velocity-space signature is identifiable, as the reduction in the integration window and the number of particles used to reconstruct distribution in Figure~\ref{fig:dhy-corr} did introduce noise to the computation of the correlation.
We are engaged in ongoing work to train a convolutional neural network on known velocity-space signatures such that we can utilize machine learning to constrain when the resolution of a particle-in-cell simulation is sufficient for this type of phase space analysis.
Of even greater interest is the use of the field-particle correlation technique to characterize the velocity-space signature, or signatures, of the instabilities which lead to the corrugation of the three-dimensional \dHy\ shock examined in Section~\ref{sec:3D}.
We will then have completely characterized the energy transfer between particles and electromagnetic fields in these particular unstable shocks, and in so doing, provide a framework for the application of this type of phase-space analysis to a broad class of collisionless shocks.
\section*{Acknowledgements}
The authors thank A. Spitkovsky for enlightening discussions on collisionless shocks. J. Juno thanks the entire \gke~team, especially A. Hakim, for all of their insights.

\section*{Funding}
This work used the Extreme Science and Engineering Discovery Environment (XSEDE), which is supported by National Science Foundation grant number ACI-1548562.
J. Juno was supported by a NSF Atmospheric and Geospace Science Postdoctoral Fellowship (Grant No. AGS-2019828) and by the U.S. Department of Energy under Contract No. DE-AC02-09CH1146 via an LDRD grant.
C. R. Brown, G. G. Howes, C. C. Haggerty, and J. M. TenBarge were supported by NASA grant 80NSSC20K1273.
G. G. Howes was also supported by NASA grants 80NSSC18K1366, 80NSSC18K1217, 80NSSC18K1371, and 80NSSC18K0643.
J. M. TenBarge was also supported by DOE grant DE-SC0020049.
K. G. Klein was supported by NASA Grant 80NSSC19K0912.

\section*{Data availability}
\gke~ is open source and can be installed by following the instructions on the \gke~ website (\url{http://gkeyll.readthedocs.io}). 
The input file for the \gke~ the simulation presented here is available in the following GitHub repository, \url{https://github.com/ammarhakim/gkyl-paper-inp}.

\section*{Declaration of interests}
The authors report no conflict of interests.

\bibliography{abbrev.bib,transverse-simulation.bib}

\appendix

\section{Complete \gke\ simulation parameters}\label{app:gke-param}
The \gke\ simulation domain extends from $[-L_x, L_x]$, $L_x = 24 d_i$, and electrons and ions are initialized with the same supersonic and super Alfvénic flow, $U_x = \pm 6 v_A$, $U_x = 6 v_A$ from $x = [-L_x,0]$ and $U_x = -6 v_A$ from $x = [0,L_x]$ such that the flows collide at $x = 0$ and produce a shockwave that propagates back towards the walls in a symmetric fashion.
We will focus on only the right half of the domain from $x = [0, L_x]$.
We use a reduced mass ratio between the ions and electrons, $m_i/m_e = 100$, the total plasma beta, $\beta = 2 \mu_0 n_0 k_B (T_{e_0} + T_{i_0})/B_0^2 = 2$, with the ion beta, $\beta_i = 1.3$, and electron beta, $\beta_e = 0.7$, and both the ions and electrons are non-relativistic, with $v_{te}/c = 1/8$, where $v_{ts} = \sqrt{2 k_B T_{s_0}/m_s}$.
Since the plasma is initialized with a flow in a background magnetic field, we initialize an electric field to support the component of the flow perpendicular to the magnetic field, $\mvec{E} = -U_x \hat{\mvec{x}} \times \mvec{B}$ which leads to a background electric field in the $y$ direction. 
We note that the initialization of the upstream distribution function with a net flow in the shock-normal direction includes a flow parallel to the magnetic field, but because the flow is parallel to the magnetic field, no further inputs are required.

The grid resolution in configuration space is $N_x = 3072$, or $N_x = 1536$ for the right half of the domain $x = [0, L_x]$, corresponding to $\Delta x \sim d_e/6 \sim 1.9 \lambda_D$, where $d_e = c/\omega_{pe}$ and $\lambda_D = v_{te}/(\sqrt{2} \omega_{pe})$ are the electron inertial length and electron Debye length, respectively.
For velocity space, the electrons are simulated on a $[-6 v_{th_e}, 6 v_{th_e}]^3$ domain with $N_v = 24^3$ grid points, and the ions are simulated on a $[-16 v_{th_i}, 16 v_{th_i}]^3$ domain with $N_v = 32^3$ grid points.
We employ piecewise quadratic Serendipity elements for the discontinuous Galerkin basis expansion \citep{Arnold:2011}.
A small collisionality is employed for regularization of velocity space, $\nu_{ee} = 0.1 \Omega_{ci}$, $\nu_{ii} = 0.01 \Omega_{ci} = e B_0/m_i$, where $\Omega_{ci}$ is the ion cyclotron frequency. 

\section{Single-particle motion of the reflected ions}\label{app:spm}
\begin{figure}
    \centering
    \includegraphics[width=\linewidth]{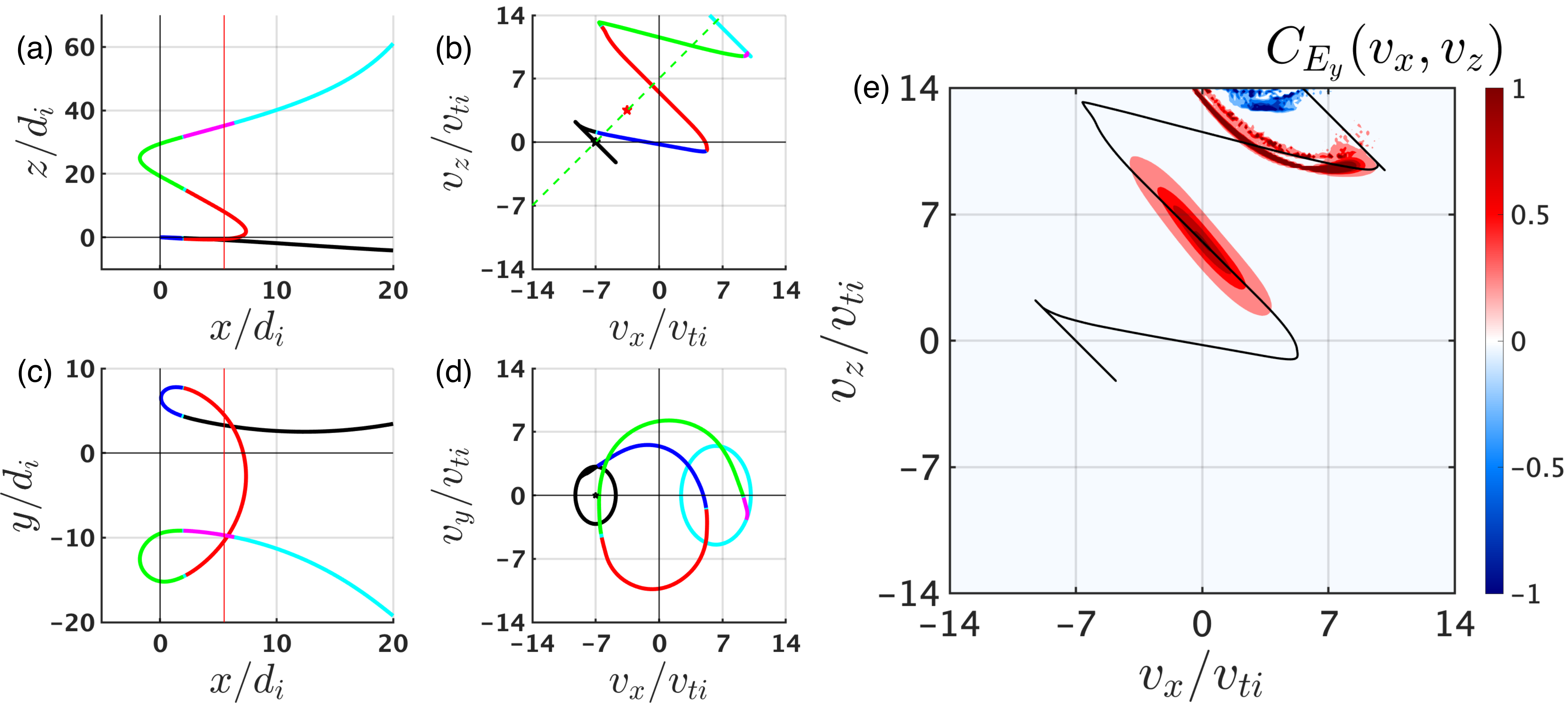}
    \caption{Particle trajectories (a)--(d) with $C_{E_y}$ field-particle correlation computed via a Vlasov-mapping technique which constructs the ion distribution function from many integrated particle trajectories (e). After the first reflection, particles are again reflected off the magnetic field gradient and can return upstream, gaining further energy along the upstream motional electric field.}\label{fig:spm}
\end{figure}

Using a similar analysis to that employed in \citet{Juno:2021}, we illustrate this multiple-bounce picture in Figure \ref{fig:spm} by combining a Lagrangian picture of single-particle motion with an Eulerian picture of the rate of energization of the full ion velocity distribution using a Vlasov-mapping technique \citep{Scudder:1986}.
We advect a number of particles in the self-consistent electromagnetic fields given by the \gke\ simulation, and then assuming the upstream distribution function is a Maxwellian and phase space is incompressible, we can reconstruct the ion distribution function through the shock.
Comparing the trajectory of a particle which has reflected twice off of the shock, we identify two diagonal features of energization in $C_{E_y}(v_x,v_z)$ coincident with the red and maroon segments of the ion trajectory.
Particles at $x = 19 d_i$, $v_x \sim [-3.5 v_{ti}, 3.5 v_{ti}], v_z \sim [0, 7 v_{ti}]$, are undergoing their first bounce and gaining energy, and $v_x \sim [-1.0 v_{ti}, 1.0 v_{ti}], v_z \sim [10, 14 v_{ti}]$ are undergoing a second bounce and gaining further energy.
The Vlasov-mapping technique allows us to connect the motion of these individual particles to the observed features in phase space when a whole population of particles are undergoing this motion and reflecting off the shock front multiple times.

\section{Complete \dHy\ simulation parameters}\label{app:dhy-param}
The full three-dimensional domain is $(L_x, L_y, L_z) = (96 d_i, 12 d_i, 12 d_i)$ with $(N_x, N_y, N_z) = (192, 48, 48)$ grid points, $\Delta x = d_i/4$.
$N_{ppc} = 10,000$ at the reference density, which we take to be the upstream density, $n_0 = 1.0$.
\dHy\ requires the ratio $v_A/c$ to be specified since the algorithm is designed to handle the generation of energetic particles self-consistently, and we pick this ratio to be small, $v_A/c = 0.008$ such that the Lorentz boost factors of the most energetic particles are still $\gamma \sim 1$.
Additionally we choose $\beta = 2 \mu_0 n_0 k_B (T_{e_0} + T_{i_0})/B_0^2 = 2$, with equal ion and electron plasma beta, $\beta_i = \beta_e = 1.0$.
Note that while total $\beta$ is the same between the \gke\ and \dHy\ simulation, the individual species values of $\beta_{e,i}$ are slightly different from the \gke\ simulation.
The electric field to support the perpendicular component of the flow is initialized identically to the \gke\ simulation.
\end{document}